\documentclass[journal=jacsau,manuscript=article]{achemso}

\usepackage{chemformula} 
\usepackage[T1]{fontenc} 
\usepackage{bm}
\usepackage{xcolor}
\usepackage{amssymb}
\usepackage{amsmath}

\newcommand\dmi[1]{{\color{black}#1}}
\newcommand\dmii[1]{{\color{black}#1}}

\author{Davide Michieletto}
\affiliation[Edinburgh]{School of Physics and Astronomy, University of Edinburgh, Peter Guthrie Tait Road, Edinburgh, EH9 3FD, UK}
\alsoaffiliation[MRC]{MRC Human Genetics Unit, Institute of Genetics and Cancer, University of Edinburgh, Edinburgh EH4 2XU, UK}
\email{davide.michieletto@ed.ac.uk}
\author{Mattia Marenda}
\affiliation[Edinburgh]{School of Physics and Astronomy, University of Edinburgh, Peter Guthrie Tait Road, Edinburgh, EH9 3FD, UK}
\alsoaffiliation[MRC]{MRC Human Genetics Unit, Institute of Genetics and Cancer, University of Edinburgh, Edinburgh EH4 2XU, UK}

\title{Rheology and Viscoelasticity of Proteins and Nucleic Acids Condensates} 

\abbreviations{IR,NMR,UV}

\begin{document}

\begin{tocentry}
\includegraphics{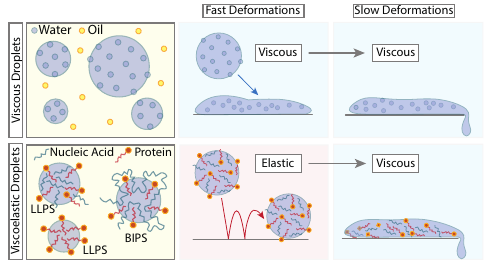}

\end{tocentry}

\begin{abstract}
   Phase separation is as familiar as watching vinegar separating from oil in vinegrette. The observation that phase separation of proteins and nucleic acids is widespread in living cells has opened an entire field of research into the biological significance and the biophysical mechanisms of phase separation and protein condensation in biology. Recent evidence indicate that certain proteins and nucleic acids condensates are not simple liquids and instead display both viscous and elastic behaviours, which in turn may have biological significance. The aim of this perspective is to review the state-of-the-art of this quickly emerging field focusing on the material and rheological properties of protein condensates. Finally, we discuss the different techniques that can be employed to quantify the viscoelasticity of condensates and highlight potential future directions and opportunities for interdisciplinary cross-talk between chemists, physicists and biologists.
\end{abstract}


\section*{Phase Separation and Rheology} 
\begin{figure*}[t!]
\includegraphics[width=1\textwidth]{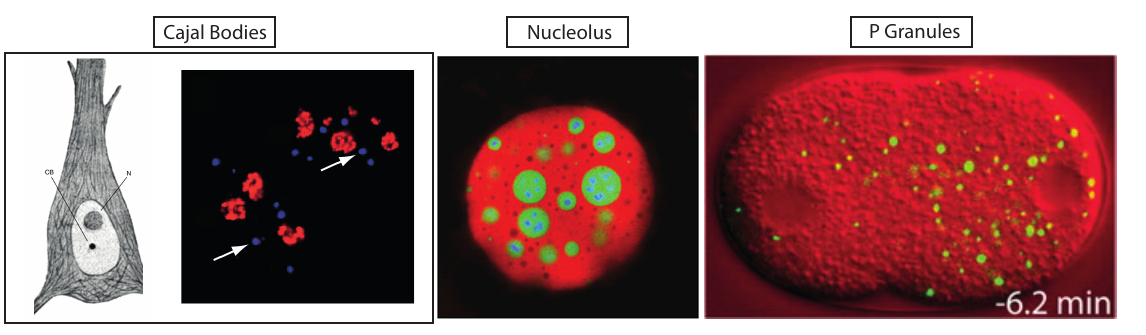}
\caption{\textbf{Examples of membraneless organelles}. From left to right: Cajal Bodies as sketched by Ramon y Cajal (reproduced with permission from reference~\cite{Matera2003}. 2003 Cell Press) and imaged in HeLa cells stained for coilin (blue) and fibrillarin (red) (this image adapted from ref.~\cite{Rino2007} is licensed under CC BY 4.0). The nucleolus of {\it X. laevis} stained for NPM1 (red), FIB1 (green) and POLR1E (blue) (reproduced with permission from reference~\cite{Feric2016}. 2016 Cell Press). Germline P Granules expressing GFP::PGL-1 (green) on differential interference contrast (red) {\it C. elegans} (reproduced with permission from reference~\cite{Brangwynne2009}. 2009 AAAS).}
\label{fig:celllphasesep}
\end{figure*}

In biology textbooks, organelles are defined as regions of space in the cell dedicated to specific operations such as the endoplastic reticulum, mitochondria or the Golgi apparatus~\cite{Alberts2014}. These physical compartments are characterised by a surrounding membrane that separates them from the rest of the cellular space. At the same time, there exist many examples of cellular compartments that are not surrounded by a membrane. The observation of membraneless compartments dates back to the early 1900 with the sketches of Ramon y Cajal discovering the eponymous bodies~\cite{ramon1903,Gall2003} (Fig.~\ref{fig:celllphasesep}a). However, the realisation of how widespread these membraneless compartments are, arrived much more recently and encompasses a range of cell bodies, from the nucleolus~\cite{Feric2016} to germline granules~\cite{Brangwynne2009} (see Fig.~\ref{fig:celllphasesep}b,c). 

It is now broadly accepted that a wide range of proteins and nucleic acids (e.g. DNA and RNA) form membraneless compartments. 
Arguably, one of the most important and open question in biology is to understand their biological significance and the biophysical mechanisms that drive their formation~\cite{Brangwynne2011,Zhu2015,Banani2017,Choi2020,Bhat2021}. Albeit existing evidence suggest that liquid-liquid phase separation (LLPS)~\cite{Alberti2021} (defined as a reversible thermodynamic process leading to the demixing of liquid fluids) is widespread and underlies the emergence of membraneless compartments, it has also been recently shown that some condensates exhibit puzzling and exotic flow behaviours and are far from being simple liquids~\cite{Shin2017,Jawerth2020,Alshareedah2021c,Ghosh2021}. As we shall discuss in detail in this review, some proteins and nucleic acids condensates display so-called viscoelastic, i.e. both viscous and elastic, flow properties akin to those of gels, foams or even rubbers~\cite{Gennes1979,Kato2012,Molliex2015a,Kroschwald2015}. These non-trivial behaviours may be due to (i) the so-called ``ageing'' of the fluid as consequence of LLPS-driven local increase in protein density~\cite{Patel2015,Jawerth2020}, (ii) the onset of percolating networks of associative ``sticker-spacer'' polymers~\cite{Alshareedah2021c,Ghosh2021,Ronceray2022,Choi2020a}, or (iii) alternative demixing mechanisms, such as bridging-induced phase separation (BIPS)~\cite{Ryu2021,Brackley2013} (see ``Models of Phase Separation'' section for a detailed discussion). In fact, while some protein condensates may display classic hallmarks of ``liquid-liquid'' phase separation such as fusion, they can either mature into solid-like structures or display subtler elastic behaviours at sub-second timescales once the high density phase is formed~\cite{Ader2010,Feric2016,Jawerth2020,Alshareedah2021c}. Thus, a condensate that originally formed by LLPS is not necessarily purely liquid at all times, and evidence of viscoelastic behaviours are increasingly more common.
Importantly, the unexpected flow behaviours observed in certain condensates are thought to be biologically relevant and intimately related to certain diseases~\cite{Shin2017a,Mathieu2020,Nozawa2020} or biological functions~\cite{Michieletto2019review}. 

To better understand the biological significance of phase separation {\it in vivo} it is therefore important to be able to quantitatively assess the material and flow properties of protein condensates.
Rheology (from ``panta rei'' or ``everything flows'' a famous quotation of Heraclitus' philosophy) is a well-established research field with strong ties to polymer physics and soft matter but perhaps less broadly known by the biological and biochemical research communities. In this review we thus aim to provide a comprehensive yet synthetic overview of concepts and techniques that can be used to quantify the rheology and viscoelasticity of protein and nucleic acids condensates with the aim of assisting the design and interpretation of existing and future experiments in this field.

\subsection*{Biophysics and Biochemistry of Phase Separation} 

\begin{figure*}
\includegraphics[width=1\textwidth]{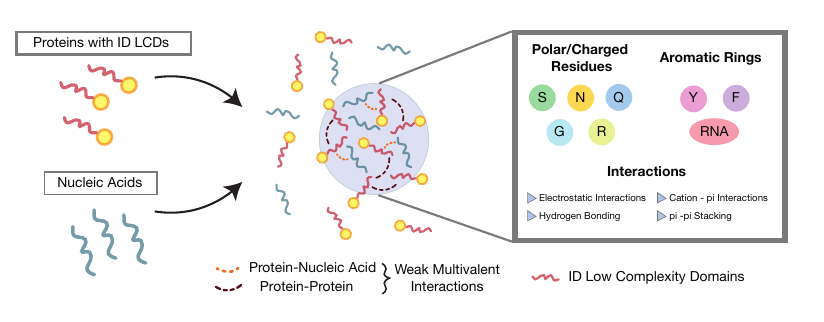}
\caption{\textbf{Biochemical interactions driving LLPS.} Proteins containing Intrinsically Disordered Low Complexity Domains (ID LCDs) and nucleic acids are prone to phase separate into membraneless condensates. The drivers for such behaviour are weak multivalent interactions, e.g. between proteins or proteins and nucleic acids, and involve polar/charged residues and aromatic rings in the protein residues and RNA. Protein-RNA interactions are typically electrostatic attractions and cation-pi interactions. Protein residues typically interact through electrostatic interactions, hydrogen bonding, cation-pi and pi-pi stacking. }
\label{fig:phasesep}
\end{figure*}

For liquids such as oil and water demixing is driven to hydrophobic interactions: it is energetically favourable for water molecules to be surrounded by other water molecules as it creates the conditions for hydrogen bond formation, in turn reducing the internal energy of the system. These interactions overcome the entropy of mixing, which would tend to keep the oil and water molecules mixed throughout the solution. The thermodynamics of this process is described at the mean field level by Flory-Huggins (FH) theory~\cite{Gennes1979,Brangwynne2015}, even in the case that long polymers such as nucleic acids are involved in the process. In the simplified FH framework, the effective interaction strength is given by the so-called ``Flory parameter'' $\chi$ which captures how favourable polymer-polymer interactions are with respect to solvent-polymer ones. When $\chi$ is larger than a critical value $\chi_c$, the system favours demixing. While the Flory parameter depends on the specific details of the system, its temperature, pH, etc., at mean field level $\chi_c$ only depends on the length of the polymers as it separates the regions where the enthalpic contributions win over the entropic ones. 

This simple picture fails for protein condensates since the details of the protein sequence matter, and thus refinement of the FH theory~\cite{Martin2018} or simulations with specific interactions~\cite{Choi2020} are necessary. The specific biochemical interactions and dependence on sequence composition that drive protein condensation, or in some cases co-condensation with nucleic acids, are not fully understood. Typically, protein condensation is driven by multi-valent weak interactions between intrinsically disordered low-complexity domains (ID LCDs) of the protein, i.e. protein segments containing significant enrichment with specific amino acids types or sequence repeats~\cite{Wootton1993,Martin2020} and that do not adopt a unique folded conformation~\cite{Romero2001,vanderLee2014}. It is typically assumed that ID LCDs that have an enrichment in polar amino acids, such as serine, asparagine, glutamine and glycine, have the potential to collapse and aggregate~\cite{Nott2015,Elbaum-Garfinkle2015,Mitrea2016,Murthy2019}. In particular, this seems more common if the strands of these polar amino acids are alternated by aromatic (tyrosine and phenylalanine) and charged (arginine) amino acids~\cite{Lin2017}. Both the patterning and the sequence position play a role in the phase separation, but few general principles have been uncovered \dmi{(for a detailed review on molecular interactions and multi-component condensates, see Ref.~\cite{Dignon2020})}.   
A specific subset of ID LCDs that has been extensively studied is the so-called RG/RGG protein domain~\cite{Chong2018b}. This is a disordered RNA-binding domain present in several nuclear proteins, such as FUS protein, and that shows repeats in arginine-glycine and arginine-glycine-glycine sequence~\cite{Thandapani2013}. The typical interactions that have been suggested to occur in RGG phase separation include electrostatic interactions, cation-pi, pi-pi and hydrogen bonding interactions~\cite{Chong2018b} (see Fig.~\ref{fig:phasesep}); in these cases, glycine and diglycine residues have an exposed peptide bond in the backbone which promotes pi-pi interactions. The same amino acids may also form the pi-pi stacking with the arginine positively charged guanidino group as well as with aromatic side chains of tyrosine and phenylalanine~\cite{Martin2018,Brady2017}. Differently from these, arginines are highly positively charged amino acids and can interact electrostatically with negatively charged or phosphorylated residues, as well as with RNA molecules phosphate groups~\cite{Altmeyer2015,Tsang2019,Boeynaems2017,Saha2016}. Interestingly, mutation of arginine to lysine on the RGG domain of Lsm4 protein has shown to impair the ability of Lsm4 to form condensed P-Bodies~\cite{Arribas-Layton2016}. The patterning of Arginine is crucial, as both experiments and simulations suggest that the distribution of charges is a key factor that determines the dynamics of phase separation and the material properties of the condensates~\cite{Lin2016b,Das2013,Nott2015}. Arginine can also promote condensation mediated by cation-pi interactions with aromatic residues, such as tyrosine and phenylalanine, and aromatic rings on RNA bases~\cite{Anbarasu2007,Zhang2014b}. The removal of aromatic residues, and in particular of tyrosine from an engineered FUS-like intrinsically disordered domain displays impaired phase separation~\cite{Lin2017}.

Besides the sequence composition of intrinsically disordered domains, condensates formation can be affected by environmental conditions such as temperature, ionic strength, pH, etc. as well as interactions between folded and disordered domains of the same protein~\cite{Alberti2017,Wu2016}. A typical example is the RNA-binding protein hnRNPA1, where the presence of folded domains reverses the salt dependence of the driving force for phase separation~\cite{Martin2021}. Another intriguing case is that of hnRNPU, an abundant nuclear protein which does not display evidence of phase separation in spite of its RNA-binding domain being an RGG repeat similar to FUS and hnRNPA1~\cite{Nozawa2017,Marenda2021b,marenda2020superstructure}. Thus, current evidence suggests that protein condensation is not only due to the intrinsically disordered domains of proteins but also how they interact with the folded, structured domains.

\section*{Viscoelasticity of Condensates}

\dmi{As opposed to water -- a so-called ``Newtonian'' fluid whose viscosity does not change as a function of the applied stress --  many protein condensates (or proteins and nucleic acids co-condensates) are not simple liquids, rather they are complex fluids with non-Newtonian behaviours~\cite{Jawerth2020,Alshareedah2021c,Ghosh2021}. Non-Newtonian fluids are characterised by the fact that their deformation rate is not trivially proportional to the amount of stress applied, or in other words, their viscosity depends on how much and how fast the sample is stressed~\cite{furst2018microrheology}. For instance, shear-thinning fluids such as shampoos, creams and ketchup display a lower viscosity when stressed and will thus flow more easily when spread over a surface. On the other end of the spectrum, shear thickening fluids such as oobleck (water and cornstarch) are more difficult to deform when quickly sheared. Other examples are yield stress materials, such as mayonnaise or shaving foam, which require a threshold stress to be attained before they start to flow at all. In general, viscoelastic fluids display viscous and elastic responses that are non-trivial functions of the timescales (or frequencies) of the perturbations at which they are subjected to. For instance, an interesting class of viscoelastic fluids called Maxwell fluids display a single relaxation timescale $\tau_R$. At deformation frequencies larger than $\tau_R^{-1}$ (fast deformations), the fluid behaves like an elastic solid, and at frequencies much smaller than $\tau_R^{-1}$ (slow deformations) it behave as a liquid~\cite{Jawerth2020,Alshareedah2021c}.}  

While LLPS is now widely argued to be involved in a range of fundamental biological processes such as gene expression~\cite{Boija2018,Sabari2018a,Cho2018,Strom2017}, the hypothesis that non-Newtonian properties of condensates may play a role in the cell biology has emerged only recently. For instance, a viscous liquid-like compartment could act as a protein reservoir or crucible to accelerate biochemical reactions~\cite{Sabari2018a,Eeftens2021}. On the contrary, elastic and gel-like RNA-protein condensates may offer local structural support to shape chromatin organisation in the nucleus~\cite{Nozawa2017,Michieletto2019rnareview}, something that cannot be achieved by purely viscous condensates. At the same time, some material properties of protein condensates may have an impact on the cell health; for example, cytoplasmic appearance of stress granules formed by heterogeneous ribonucleoproteins such as hnRNPA1 and solid-like elastic fibrous structures are commonly observed in degenerative diseases such as amyotrophic lateral sclerosis, suggesting that the onset of solid-like properties of these condensates may be linked to the onset of the disease~\cite{Molliex2015a,Shin2017a}. 

\dmi{Quantitative studies on the viscoelasticity of protein condensates has started only very recently~\cite{Jawerth2018,Jawerth2020,Alshareedah2021c,Ghosh2021}. To provide the reader with an overview of the typical values of viscous and elastic properties, we report a list of different proteins and nucleic acids condensates with their values of, where available, viscosity, surface tension and elasticity (see Table~\ref{tab:table}).}
\begin{table}[t!]
    \centering
\begin{tabular}{c|c|c|c|c|c}
Protein/body & $\eta$ [Pa s] & $\gamma$ [$\mu$N/m] & $G^{\prime}_p$ [Pa] & Probes/Method & Refs.\\
\hline 
Human Nucleus & 0.001-0.003 & -- & -- & GFP FCS & \cite{Baum2014} \\
Mouse Nucleus & 25.1 & -- & 0.48 & MR, 200 nm nanorods & \cite{Celedon2011} \\
Mouse Nucleus & 52 & -- & 18 & MR, 100 nm & \cite{Tseng2004} \\
Human Nucleus & 1200 & -- & 250 & MR, 1 $\mu$m & \cite{DeVries2007}\\
Human Nucleus & 3000 & -- & -- & shape fluctuations & \cite{Caragine2018} \\
Human Nucleolus & -- & 1 & -- & shape fluctuation & \cite{Caragine2018} \\
{\it X. Laevis} Nucleolus & 12-32 & 0.4 & -- & coalescence & \cite{Feric2016} \\
NPM1 & 0.74 & -- & no & MR, 50 nm & \cite{Feric2016} \\
\textit{C. Elegans} P granules  & 1 & 1 & -- & coalescence & \cite{Brangwynne2009} \\
TDP-43 & 0.01-3.7 & -- & -- & coalescence & \cite{Gopal2017} \\
LAF-1 & 8-34 & 100 & -- & coalescence/MR, 100 nm & \cite{Elbaum-Garfinkle2015} \\
LAF-1 RGG & 1.62 & 159 & -- & micropipette aspiration & \cite{Wang2021} \\
PGL-3 (75 mM KCl) & 1 & 5 & 15 & active MR, 1$\mu$m & \cite{Jawerth2018} \\ 
PGL-3 (180 mM KCl) & 0.1 & 1 & 0.1 & active MR, 1$\mu$m & \cite{Jawerth2018} \\ 
PGL-3 (early, 75mM KCl) & 4.4 & 4.5 & 56 & active MR, 1$\mu$m & \cite{Jawerth2020} \\ 
PGL-3 (late, 75mM KCl) & 40 & 19.3 & 50 & active MR, 1$\mu$m & \cite{Jawerth2020} \\ 
FUS (early) & 4 & -- & 0.4 & active MR, 1$\mu$m & \cite{Jawerth2020} \\
FUS (late) & 50 & -- & 0.1 & active MR, 1$\mu$m & \cite{Jawerth2020} \\
FUS & 0.01-0.1 & -- & -- & FRAP in vivo & \cite{Patel2015b} \\
FUS & 1 & 100 & no & sessile drop/MR, 100 nm & \cite{Ijavi2021} \\
$\left[\textrm{KGKGG}\right]_5-\textrm{rU}_{40}$ & 0.26 & -- & no & passive MR, 200 nm \& 1$\mu$m & \cite{Alshareedah2021c} \\
$\left[\textrm{RGPGG}\right]_5-\textrm{rU}_{40}$ & 0.1 & -- & no &  passive MR, 200 nm \& 1$\mu$m  & \cite{Alshareedah2021c} \\
$\left[\textrm{RGRGG}\right]_5-\textrm{rU}_{40}$ & 6 & -- & 60 &  passive MR, 200 nm \& 1$\mu$m & \cite{Alshareedah2021c} \\
water & 0.001 & 72000 & no & -- & -- \\
honey & 10 & 50000 & 10-100 & -- & -- \\
\end{tabular}
\caption{Table of viscosity $\eta$, surface tension $\gamma$ and elasticity $G^\prime_p$ (extracted from the maximum frequency that could be measured in the cited work) for different natural and synthetic protein condensates. Generic values for water and honey are given as reference. MR stands for microrheology. Unless specified, the probes for microrheology are spherical particles of given size. The $G_p^\prime$ table entry is marked as ``no'' if no elasticity was observed and otherwise ``--'' if not measured.}
\label{tab:table}
\end{table}
\dmi{The table should also give the reader a sense of the heterogeneity in the values measured for similar condensates in the literature. As we shall explain in detail in the next section, these measurements are sensitive to the technique and probes employed. For instance, while small particles and GFP molecules are more suited to be embedded into cells and cell nuclei, they may be smaller than the pore or mesh size of their surrounding environment -- around 10-100 nm for chromatin~\cite{Gorisch2005,Baum2014} compared with $\lesssim$ 10 nm of a GFP molecule -- and therefore may not fully capture the viscosity and elasticity of the bulk environment. A glaring example is the apparent viscosity of the nucleus, found to be comparable to that of water by performing FCS on GFP molecules~\cite{Baum2014} and 3 million times larger using nuclear shape fluctuations~\cite{Caragine2018}. In the next section, we will describe the different techniques that can be used to measure the viscoelastic behaviour of protein and nucleic acids condensates.}

\subsection{Microrheology} 
Classical bulk rheology is typically performed on large samples by placing $\sim$ml of sample in between plates that are made to rotate relative to each other with chosen amplitude and frequency so to shear the sample. By measuring the force experienced by the plates one can estimate the viscous and elastic components of the material as a function of amplitude of the strain and shear rate. While bulk rheology is widely employed in industrial settings, protein condensates are not amenable to this technique because they (i) typically appear in droplets within other fluids and (ii) they are often difficult to produce at milliliters scale. 

One popular choice to measure the viscoelasticity of scarce samples is active or passive ``microrheology''~\cite{Mason1995}, a method that employs spherical particles embedded in the fluid of interest to probe its material properties. In the last 10-20 years, microrheology has been extensively used to characterise cellular structures {\it in vitro} as well as {\it in vivo} such as the cytoplasm, cytoskeleton, nucleoplasm, etc.~\cite{Kasza2007,Wirtz2009,Hameed2012}. 
\dmi{Microrheology can be done in passive or active mode. Passive microrheology leverages the thermal diffusion of particles within the fluid to extract information on its material properties~\cite{Wirtz2009}. A limitation of this technique is that it can only probe small deformations of the sample, driven by thermal noise alone. On the other hand, active microrheology employs optical tweezers to apply larger-than-thermal forces on the beads embedded in the fluid and measures the response of the fluid~\cite{Jawerth2018,Fitzpatrick2018}. Note that there are techniques sitting in between the two and that use optical tweezers to trap the particles in place and study their thermal fluctuations~\cite{Tassieri2012,Evans2009}.}

\dmi{Using microrheology Jawerth et al~\cite{Jawerth2018,Jawerth2020}, Alshareedah et al~\cite{Alshareedah2021c} and Ghosh et al~\cite{Ghosh2021} reported the most quantitative studies on the viscoelasticity of condensates {\it in vitro} so far. Jawerth and co-authors studied protein droplets formed by PGL-3 and proteins of the FUS family (FUS, EWSR1, DAZAP1, and TAF15) and found that these condensates behaved as ``ageing'' Maxwell fluids, i.e. as fluids displaying a frequency-dependent viscoelastic response with a relaxation timescale that became longer over time (see below). An intriguing discovery of their work is that the elastic plateau -- i.e. the measure of elastic solid-like response of the condensate -- does not appear to increase over time in ageing droplets: older condensates are not harder than their younger counterparts. At the same time, this aging, or maturation, behaviour was found to increase the intrinsic relaxation time of the fluid over time. This may reflect an internal rearrangement dynamics of the protein condensate; the macromolecular components re-organise to find deeper energy minima within the large configurational space in turn increasing the timescales of elastic response of the condensate. As pointed out in Ref.~\cite{Jawerth2020}, this mechanisms is not dissimilar to the physics of certain glasses and from a biological standpoint, this aging process may eventually lead to pathological and irreversible condensates akin to amyloid fibrils.}
\dmi{At the same time, Alshareedah et al~\cite{Alshareedah2021c} performed passive microrheology on optically trapped beads (see below) and discovered that sequence composition of short poly-peptides has a marked impact on the material properties of protein-RNA condensates. For example, they find that the condensates behave as Maxwell fluids, and that poly-peptides with sequences $[$RGXGG$]_5$ with X=$\{$P,S,R,F,Y$\}$ display increasing values of viscosity from $\sim$ 0.1 Pa s to $\sim$ 40 Pa s. They also find that the relaxation time $\tau_R$ determining the timescale of elastic response varies from 0 (for P,S residues) to 1 second for $[$RGYGG$]_5$ for which the elastic plateau reaches up to 60 Pa. They also find that the presence and sequence of RNA has an effect on the condensate material properties and that the viscoelastic behavior is correlated with the strength of protein-nucleic acid interactions. Finally, Ghosh et al~\cite{Ghosh2021} used active microrheology (see below) to show that heterotypic protein condensates form viscoelastic fluids and that the timescales of fusion in coalescence experiments (see below) is governed by the elastic component of the condensate. These studies highlight the impact that single residue mutations can have on the behaviour of the condensates and the ability of microrheology to provide quantitative and precise information on the flow properties of condensates. We thus now describe how to perform and critically interpret passive and active microrheology techniques in detail. }

\subsubsection{Passive Microrheology} 
\begin{figure*}[t!]
\includegraphics[width=1\textwidth]{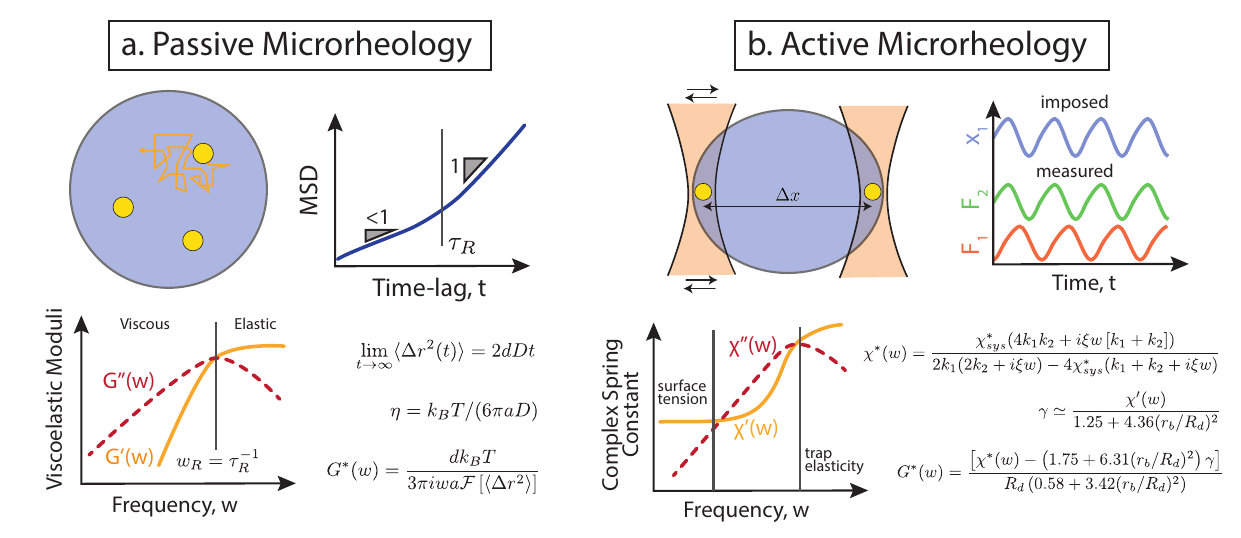}
\caption{\textbf{Microrheology of condensates.} \textbf{a.} Passive microrheology is performed by tracking the position of probe particles within a condensate. From the mean squared displacement, $\langle \Delta r^2(t) \rangle$, one can obtain the elastic and viscous moduli $G^\prime$, $G^{\prime \prime}$ which in turn quantify the full viscoelastic spectrum of the condensate and its viscous and elastic components as a function of deformation frequency. \textbf{b.} Active microrheology is performed by trapping beads within a condensate and using them as ``halndes'' to apply an oscillatory stress to the droplet. The complex and frequency dependent spring constant of the droplet can be computed by measuring the distance of, and forces experienced by, the particles as a function of the frequency of oscillations.}
\label{fig:panel_urheo}
\end{figure*}

In a typical passive microrheology experiment, a brightfield or epifluoresce microscope is used to record movies of passive spherical particles (typically around 0.1-1 $\mu$m in size) diffusing within the condensate. From these movies, particle tracking algorithms (such as TrackPy~\cite{Crocker1995} or TrackMate~\cite{Tinevez2017} in ImageJ) are used to obtain the trajectories of the particles, $\bm{r}(\tau)$ (Fig.~\ref{fig:panel_urheo}a). In turn, the mean squared displacement (MSD) for a given (lag-)time $t$ is computed from the trajectories as 
\begin{equation}
\langle \Delta r^2(t) \rangle = \langle \left[\bm{r}(\tau+t) - \bm{r}(\tau)\right]^2\rangle \, ,
\label{eq:MSD}
\end{equation} 
\dmi{where the average is intended over particles and times $\tau$. In other words, the mean squared displacement measures the (square) length explored by a particle in between any two points of its trajectory within a given time window $t$. For example, a fast particle may cover $1 \mu$m$^2$ in $t = 1$ second, while a slow one may explore $0.01 \mu$m$^2$ in the same $t=1$ second. If we assume that the sample around the particle is at steady state, we expect that the particle mobility will not change in time during the observation. This implies that we expect the particle to have the same mobility in the first and last seconds of its trajectory, and for this reason we can take the average over initial times $\tau$ in Eq.~\eqref{eq:MSD}. In turn, the MSD is thus only a function of {\it lag-time} $t$. Importantly, for aging systems, such as FUS or PGL-3 droplets~\cite{Jawerth2020}, one should take care that the temporal average is shorter than the timescales over which the fluid changes its material properties to avoid confounding effects due to the particle mobility changing over the observational timescale.}

In a viscoelastic fluid, being it a condensate or not, the behaviour of the MSD at \emph{large enough times and unconfined space} is described by $\lim_{t \to \infty} \langle \Delta r^2 (t) \rangle = 2d D t$, i.e. a freely diffusive behaviour with diffusion coefficient D and $d=1, 2$ or $3$ being the dimensionality of the tracked trajectory $\bm{r}(\tau)$. For passive spherical tracers, the Stokes-Einstein relation connects the diffusion coefficient of the tracers $D$ to the viscosity of the surrounding fluid $\eta$ as 
\begin{equation}
\eta = \dfrac{k_B T}{6 \pi D a }
\label{eq:SEeq}
\end{equation}
with $a$ being the radius of the tracer, $k_B$ the Boltzmann constant and $T$ the system temperature. In some experimental conditions, and especially {\it in vivo}, it may not be possible to track the tracer beads for long enough times to observe their freely diffusive behaviour, for instance because of droplet movements that cannot be corrected or because the tracers sediment to the bottom of the sample. In these cases, one has to rely on measurements at shorter timescales, where the MSD may assume a more generic functional form 
\begin{equation}
    \textrm{MSD}(t) = K t^{\alpha}
\end{equation}
where $\alpha$ is the exponent that describes whether the motion is constrained (also called sub-diffusive, $\alpha<1$) or active (also called super-diffusive, $\alpha > 1$). \dmi{Importantly $K$ is not a diffusion constant but rather a generic transport coefficient with units of length$^2$/time$^{\alpha}$ and if $\alpha < 1$ it cannot be used to extract the viscosity of a fluid via the Stokes-Einstein equation. In fact, in this case Eq.~\eqref{eq:SEeq} would not even have the correct units of viscosity. }

The conceptual leap to connect the tracers' trajectories to the frequency-dependent material properties of the condensate is done by realising~\cite{Mason1995} that the MSD in the time-domain is connected to the so-called complex modulus in the frequency-domain $G^*(w)$ via a unilateral Fourier transform (or a Laplace transform followed by an analytic continuation) also known as Generalised Stokes-Einstein Relation (GSER)~\cite{Mason1995,Mason2000}
\begin{equation}
G^{*}(w) = G^\prime(w) + i G^{\prime \prime}(w)= \dfrac{d k_BT}{3 \pi i a w \mathcal{F}[\langle \Delta r^2 (t) \rangle]}
\label{eq:gser}
\end{equation}
where $d=1,2$ or $3$ is the dimension of the position vector used to compute the MSD and $\mathcal{F}[\cdot]$ indicates the unilateral Fourier transform. 
The complex modulus $G^*$ is a powerful function that describes the material properties of the complex fluid in the frequency or time domains. From the complex modulus, the elastic and viscous moduli $G^{\prime}$ and $G^{\prime \prime}$, respectively, can be obtained as the imaginary and real parts of $G^{*}$. These frequency-dependent functions encode the propensity of the complex fluid to react like a viscous fluid or as an elastic solid to a linear deformation with frequency $w$. One of the simplest models of viscoelastic fluids is the so-called Maxwell fluid, which displays a crossover frequency $w_R$ at which the fluid switches from viscous to elastic behaviour (see Fig.~\ref{fig:panel}a). For frequencies $w<w_R$, i.e. timescales $t>\tau_R=w_R^{-1}$, the material flows like a liquid and this is reflected by the fact that $G^{\prime \prime}> G^{\prime}$; on the contrary, at frequencies $w>w_R$, or timescales $t<\tau_R$, fast deformations trigger an elastic response, i.e. $G^{\prime \prime} < G^{\prime}$. 

In the context of microrheology, it is also worth mentioning the importance of accurately measuring the ``noise floor''~\cite{Savin2005}. Indeed, the noise from experimental conditions, equipment, etc. will affect the precision at which particles are tracked and will in fact appear as a non-zero plateau in the MSD curves at early times. In turn, this will be transformed by the GSER into an elastic contribution of $G^\prime(w)$ at large frequencies. Thus, the noise must be assessed by measuring the MSD of beads immobilised on the cover slip and under the same experimental conditions as the other tracers. Failing to do so may yield an incorrect interpretation of the results and an overestimation of the elasticity of the condensate.

We note that passive microrheology stands out from other techniques as it is minimally invasive. For instance, bulk rheology shears the sample by pressing on it, in turn inducing some stiffening~\cite{Ciccone2020}. At the same time, it can only probe small deformations of the fluid as the motion of the beads is thermally driven. Arguably, the forces experienced by condensates {\it in vivo} may be larger than thermal ones and in the next section we discuss how these ``non-linear'' regimes may be explored.   

\dmi{Finally, we mention that particle tracking is not the only way to obtain MSD curves from the recording of particles diffusing in a system. In fact, one can also employ Differential Dynamic Microscopy (DDM), a technique that relies on the spatial and temporal autocorrelation of the pixel intensities in order to extract the dynamics of the particles in the system~\cite{Martinez2012,Cerbino2008,Edera2017}. This technique is particularly well suited for particles that are too small to be resolved with optical microscopy or to measure the dynamics of fluorescently labelled molecules such as small DNA plasmids~\cite{Smrek2021}.}

\dmi{\subsubsection{Optical Trap Microrheology} 
In microrheology, the longer the particles are imaged for the better the average and the broader the spectrum of frequencies that can be sampled. Yet, in some experimental conditions it may be difficult to image the particle for long times, because of sedimentation or other experimental challenges. For this reason it may be more appropriate to trap the particles using an optical tweezer and measure their thermally induced displacements~\cite{Alshareedah2021c,Ghosh2021}. The optical tweezer effectively traps the particle in a harmonic potential with certain stiffness, that should be calibrated and that can be tuned by setting the laser power. Because of the trap, the particle is not allowed to escape and freely diffuse in the sample. Its diffusion is thus constrained at large times and the (squared) displacements will display a plateau that is related to the trap stiffness $\kappa$ as $\langle r^2 \rangle = 3 k_BT/\kappa$ where $\langle r^2 \rangle$ is the time-independent variance of the displacement vector $\bm{r}(t)$ or, in other words, is the value of the MSD at infinite time. Using the {\it normalised} mean squared displacement (NMSD) $\langle \Delta r^2(t) \rangle_n \equiv \langle \Delta r^2(t) \rangle/2 \langle r^2 \rangle$ one can obtain the complex modulus as~\cite{Tassieri2012}  
\begin{equation}
G^*(w) = G^\prime(w) + i G^{\prime \prime}(w) = \dfrac{\kappa}{6 \pi a} \left( \dfrac{1 - i w \langle \tilde{\Delta r^2}(w) \rangle_n }{i w \langle \tilde{\Delta r^2} (w)\rangle_n}\right) \, ,
\end{equation}
where $\langle \tilde{\Delta r^2} (w)\rangle_n$ is the Fourier transform of the NMSD. Several post-processing, oversampling and optimisation protocols have been developed for this technique and it can give up to 5-6 decades of viscoelastic spectrum with a single measure~\cite{Tassieri2012,Evans2009}.}  

\dmi{\subsubsection{Active Microrheology} 
Active microrheology, where a bead is trapped and moved around the sample by an optical trap, is widely employed to probe the so-called ``non-linear'' response of a system to larger-than-thermal forces~\cite{Chapman2014prl,Fitzpatrick2018}. Additionally, this technique is appropriate for systems that are simply too stiff to be studied by thermal fluctuations only and in which the passive tracers hardly move at all. Recently, a novel active microrheology technique has been developed to extract the complex moduli and surface tension from protein droplets. The technique relies on two beads embedded in a condensate and trapped by two independent optical tweezers. One of the tweezers is kept static while the other is made to move thereby creating an oscillatory stress on the sample~\cite{Jawerth2018,Ghosh2021} (Fig.~\ref{fig:panel_urheo}b). During the experiment, the forces acting on both beads are measured and used to determine the frequency-dependent effective spring constant of the whole system (droplet and traps) as 
\begin{equation}
    \chi_s^*(w) = \dfrac{\tilde{F}_2 - \tilde{F}_1}{2 \tilde{\Delta x}} \, , 
\end{equation}
where $\tilde{F}_i$ is the Fourier transform of the force on bead $i$ and $\tilde{\Delta x}$ the Fourier transform of the relative position of the beads $\Delta x(t)$. One can then derive the complex spring constant encoding only the viscoelasticity of the droplet as~\cite{Jawerth2018}
\begin{equation}
    \chi^*(w) = \chi^\prime(w) + i \chi^{\prime \prime}(w) = \dfrac{\chi_s^* \left[ 4 k_1 k_2 + i \xi w (k_1 + k_2)\right]}{2 k_1 (2 k_2 + i \xi w) - 4 \chi_s^*(k_1 + k_2 + i \xi w)} \, . 
\end{equation}
From this equation, it is possible to extract two contributions: first, the surface tension, which dominates at slow deformations, i.e. small $w$, as  
\begin{equation}
    \gamma \approx \dfrac{\chi^\prime(w)}{1.25 + 4.36 (r_b/R_d)^2}
\end{equation}
which is valid in the limit that the bead is small compared with the droplet ($r_b \ll R_d$) and, second, the full complex modulus as
\begin{equation}
    G^*(w) \approx \dfrac{\left[ \chi^*(w) - (1.75 + 6.13 (r_b/R_d)^2)\right]\gamma }{R_d (0.58 + 3.42 (r_b/R_d)^2} \, ,
\end{equation}
which is again valid if $r_b \ll R_d$ and if $|G^*(w)| R_d \gg \gamma$.
}
  
\dmi{While this is a useful and quantitative technique, it requires a microscopy set up with two independent and finely calibrated optical tweezers which is not common in molecular biology laboratories. It also requires the production of large droplets ($\sim$10 - 20 $\mu$m) and spherical particles with large refractive index that can be trapped inside the condensate. Additionally, while it can yield both the surface tension and the viscoelastic spectrum in one measurement, the range of usable frequencies for $G^*(w)$ is relatively small, given the fact that the small ($w \lesssim 0.1$ s$^{-1}$) and large ($w \gtrsim 10$ s$^{-1}$) regimes are dominated by surface tension and trap stiffness}. 

\dmi{ Overall, we find that passive and active microrheology are currently the best techniques to quantify in full the viscous and elastic properties of condensates. Indeed, other techniques such as FRAP and FCS (see below) cannot yield a full viscoelastic spectrum of the condensate. In spite of this, microrheology is not a common technique in biology (yet): active microrheology would be very challenging to perform {\it in vivo}, on the other hand passive microrheology has been performed {\it in vivo}~\cite{Wirtz2009} but it requires microinjection to be performed in the nucleus~\cite{DeVries2007,Hameed2012}. Below, we thus describe more widely employed techniques in biology which can also give information on the condensates material properties albeit not as complete as microrheology. }

\dmii{Finally, we note that it is possible to measure viscosity and surface tension of {\it in vitro} condensates bypassing microscopy or optical tweezers by using ``micropipette aspiration''~\cite{Wang2021}}. 

\subsection{Fluorescence based techniques} 

\subsubsection{Fluorescence Recovery After Photobleaching} 

\begin{figure*}
\includegraphics[width=1\textwidth]{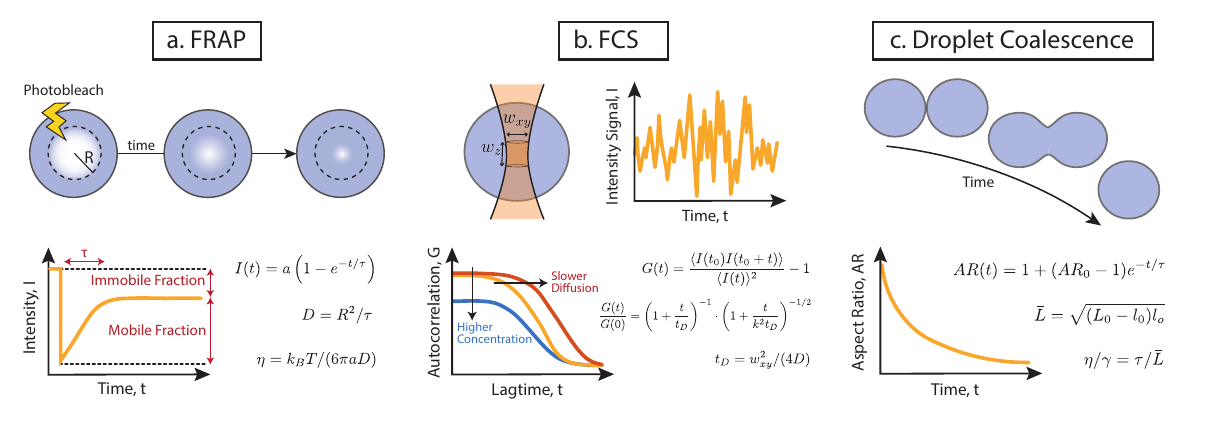}
\caption{\textbf{Fluorescence based methods to characterise viscosity and surface tension of condensates.} \textbf{a.} FRAP is performed by bleaching a region (disk, strip, whole) of the condensate. The FRAP recovery curves are typically fitted by a single expoenential function which returns a single characteristic timescale of protein rearrangement within the droplet. This timescale can be translated into an apparent diffusion coefficient of the proteins $D = R^2/\tau$~\cite{Elbaum-Garfinkle2015} which in turn may be used to crudely estimate the apparent viscosity $\eta$ via the Stokes Einstein relation~\cite{Brangwynne2009}. \textbf{b.} FCS is performed by detecting fluorescent molecules (typically GFP or proteins) diffusing through a small ($\sim$ femtoliter) volume within the sample. After fitting the autocorrelation function $G(t)$ the apparent diffusion constant is obtained as $D = w_{xy}^2/(4\tau)$, where $w$ is the size of the sampled volume. A big advantage of FCS over FRAP is that it can quantitatively measure the  concentration of molecules in the illuminated volume. \textbf{c.} Passive droplet coalescence can quantify the relaxation timescale of droplets. Assuming pure Newtonian behaviour, the balance of viscosity and surface tension determines the typical relaxation timescale as $\tau = \eta/\gamma L$. Active droplet coalescence, done by pushing together droplets trapped by optical tweezers, can overcome some technical problems of passive coalescence experiments. }
\label{fig:panel}
\end{figure*}

Fluorescence Recovery After Photobleaching (FRAP) is a popular technique to study the dynamics of cell components and has now been extensively employed to measure the dynamics of phase condensates. FRAP requires fluorescently-tagged proteins and strong lasers and is thus typically performed on a confocal microscope. It works by permanently inactivating (``bleaching'') some of the fluorescent proteins in the sample within a region of the cell or droplet, which could be a disk of radius $R$, a strip, or a larger region (see Fig.~\ref{fig:panel}a). After the bleaching step, the intensity of the fluorescence signal within the bleached region (and that in a control region) are monitored over time. Hence, this technique probes the dynamical rearrangements of the macromolecular components of the cell or droplet either within the droplet or with the soluble pool.  
Typical FRAP curves display a sudden drop (the bleaching step) followed by a recovery of the normalised intensity signal. \dmi{This recovery mirrors the diffusion, binding and reaction activity of the fluorescently labelled molecules or proteins within the sample; a process that can be modelled as a system of differential reaction/diffusion/advection equations with single or multiple binding states~\cite{Soumpasis1983,Axelrod1976,Mueller2008}. It has been noted that depending on the model used to fit these curves the values obtained for the on and off binding rates of certain transcription factors can vary widely~\cite{Mueller2008,Mazza2012a}. The analysis and interpretation of these curves is thus a critical step. In absence of an {\it a priori} hypothesis on which model to use, the most agnostic way to extract information from FRAP curves is to numerically estimate the half time (time at which the signal has recovered half of its original value) and the intensity of the long time plateau~\cite{Taylor2019}. Alternatively, for simple reaction processes one may also fit these FRAP recovery curves with a function of the form~\cite{Taylor2019}
\begin{equation}
I(t) = a(1-e^{-t/\tau})
\label{eq:frap}
\end{equation}
where $\tau$ is the mean recovery time and $a$ the fraction of mobile proteins, i.e. the part of the signal that is recovered at infinitely large times (see Fig.~\ref{fig:panel}a). The mean recovery time $\tau$ is a characteristic time of reorganisation within the cell or droplet for the fluorescently-tagged protein. From it, one can extract a characteristic diffusion coefficient as $D = R^2/\tau$, with $R$ the size of the bleached region, and in turn obtain an apparent viscosity from the Stokes Einstein equation $\eta=k_BT/(6 \pi a D)$, where $a$ is here the expected radius of the protein of interest (see Fig.~\ref{fig:panel}a). This reasoning assumes that the droplet is a simple Newtonian fluid and cannot distinguish binding and diffusion of the protein of interest. For instance, if the recovery time is very long, it cannot dissect the reason is because the protein has long binding times or slow diffusion. Albeit crude, this approach has been extensively used in the literature to estimate the viscosity of protein droplets. For instance, by performing FRAP on large {\it C. elegans} P granules {\it in vivo}, their viscosity was estimated to be~\cite{Brangwynne2009} $\simeq 1$ Pa s, which was broadly consistent with the viscosity measured via microrheology on the P granules component LAF-1 {\it in vitro} and in presence of RNA~\cite{Elbaum-Garfinkle2015b}.} 

\dmi{Another information that can be extracted from FRAP curves without making any {\it a priori} hypothesis on the dynamics and binding modes of the protein of interest is the large time recovered fraction. In Eq.~\eqref{eq:frap}, the coefficient $a$ represents the recovered fraction of the signal while its counterpart $1-a$ is the fraction of signal that has not recovered. This is typically referred to as the ``immobile'' fraction and reflects an intrinsic solid-like behaviour of the protein of interest in the bleached region. For this reason, measuring the mobile and immobile fractions can qualitatively establish the liquid or solid nature of the droplets. For instance, incomplete FRAP recovery was interpreted as a qualitative indication of viscoelasticity of fibrillarin droplets~\cite{Feric2016}. }

\dmi{FRAP is widely employed because it is a technically straightforward technique that can be performed with a confocal microscope and yields a quick readout. If the scope is to extract qualitative information on the sample, then FRAP does not require complex post-processing analysis and Eq.~\eqref{eq:frap} can be used to fit the curves~\cite{Taylor2019}. On the other hand, if the scope is to extract quantitative and precise information on binding and diffusion modes of proteins, then FRAP requires more complicated analysis and fitting models~\cite{Mueller2008}. }

\dmi{An additionaly advantage of FRAP is that it is suitable {\it in vivo} as it only requires that cells express a fluorescently labelled protein. A potential shortcoming in this respect is that overexpressing a protein {\it in vivo} may saturate its binding sites or compete with the endogenous protein species in turn generating confounding results when compared with knock-ins or single-particle tracking~\cite{Mazza2012a}. At the same time, the fluorescent tags that are typically used are, in some cases, as big as the proteins themselves thereby affecting their native state and dynamics. For example, while it is common practice to fuse a GFP tag on the protein of interest, this may itself interfere with the correct protein function. Other less invasive labelling methods may thus be preferred~\cite{Milles2012}. Additionally, proteins are typically smaller than the pore size of the surrounding mesh and hence their apparent diffusion coefficient likely underestimates the bulk viscosity of the system~\cite{Baum2014}.} 

\dmi{Another important source of potential confusion in FRAP results is that in heterotypic condensates recovery curves may be widely different depending on which population of proteins are considered as fluorescent probes. This is due to the fact that different proteins and/or nucleic acids in a condensate may display different structural roles with shorter/longer relaxation times. For instance, condensates of HP1$\alpha$ with DNA fragments display recovery curves that are slower for longer DNA fragments if DNA is used as fluorescent probe; on the contrary, no change in recovery time is observed if the fluorescently labelled probe is  HP1$\alpha$~\cite{Keenen2021}. Similar results were seen in DNA condensates mediated by crowders or H1~\cite{Muzzopappa2021}. This indicates that these condensates are viscoelastic, with the DNA providing the elastic contribution. These observations can be well explained by the ``bridging induced phase separation'' (BIPS) model, explained below. Hence, in case of heterotypic condensates fast dynamics of one of the components may be erroneously interpreted as indicating a purely liquid droplet and FRAP should thus be performed on all the components of a condensate to estimate its bulk viscoelastic nature.}

\subsubsection{Fluorescence Correlation Spectroscopy} 
\dmi{A popular high-resolution technique that is even more powerful than FRAP is Fluorescence Correlation Spectroscopy (FCS). It employs a confocal microscope to illuminate a $\sim$femtoliter volume in the sample (see Fig.~\ref{fig:panel}b). Fluorescently tagged proteins travel through the illuminated volume and fast detectors are used to record the variation in the intensity of the signal. The time trace of the signal is acquired and auto-correlated as~\cite{Langowski2008}
\begin{equation}
G(t) = \dfrac{\langle I(t_0) I(t_0+t) \rangle}{\langle I(t) \rangle^2} -1 \, ,
\end{equation}
where the average is intended over times $t_0$ and the $-1$ is there to ensure that $G(\infty) = 0$. Even neglecting sub-microsecond correlations (not relevant for this review), the analysis and fitting of FCS autocorrelation curves can be as complicated as the ones for FRAP curves. Different fitting models with multicomponent diffusion, advection and reaction kinetics have been proposed~\cite{Elson1974,Tian2011,Yu2021}.  
In the simplest scenario, with one diffusing component, the autocorrelation curve takes the functional form 
\begin{equation}
    G(t) = G(0) \left(  1 + \dfrac{t}{t_D}\right)^{-1} \left( 1+ \dfrac{t}{k^2 t_D}\right)^{-1/2}
\end{equation}
where $G(0) = \pi^{3/2}w_{xy}^2 w_z/c$ is related to the average concentration of probes $c$ in the illuminated volume of sizes $w_{xy}$ and $w_z$ ($k$ is the ratio $k = w_z/w_{xy}$) and $t_D = w_{xy}^2/(4D)$ is the time it takes for a probe to diffuse through the illuminated volume. This fitting model with a single diffusing component was used to extract the relaxation time $t_D$ (and hence apparent diffusion) of protein components in optogenetic droplets~\cite{Bracha2018a} but multiple diffusing populations can also be used if needed. As in the case of FRAP, the characteristic timescale of the protein diffusion can be translated into an apparent diffusion constant and, in turn, to an estimate of the solution viscosity, provided that the small size of protein probe is acknowledged~\cite{Alshareedah2021b,Martin2020,Bracha2018a,Alshareedah2020a,Maharana2018}. The big advantage of using FCS over FRAP is that beyond measuring protein dynamics, it can also measure protein concentration which is extremely useful in order to compile a quantitative phase diagram of the phase separated system~\cite{Bracha2018a,Martin2020}. }

\dmi{We mention that while performing non-conventional FCS methods such as multiscale FCS~\cite{Baum2014} and raster image correlation spectroscopy~\cite{DiRienzo2014} can give effective mean squared displacements of the probes by measuring spatial and temporal intensity fluctuations (akin to differential dynamic microscopy~\cite{Martinez2012}) they cannot be translated into the complex modulus $G^*(w)$ via the generalised Stokes Einstein Relation (Eq.~\eqref{eq:gser}) because it assumes that the probes are bigger than the pore size of the surrounding fluid~\cite{Mason1995}. On the contrary, GFP molecules and fluorescently-tagged proteins are typically smaller than pore or mesh size of their surrounding environment. For instance, compare the $\sim$10-100 nm for chromatin~\cite{Gorisch2005,Baum2014} with the typical size of GFP $\sim$5 nm. In turn, this implies that the MSDs obtained from FCS give information on the ``nanorheology'' of the system and are likely to underestimate the mesoscale viscosity and elasticity of the bulk.}

\dmi{We finally mention that alongside FRAP and FCS, single particle tracking using, for example, photoconvertible dyes and/or super-resolution techniques are becoming widely employed and precious tools to obtain high-resolution information on the dynamics of proteins in these condensates~\cite{McSwiggen2019a,Izeddin2014,Cho2016,Cho2018b,Hilbert2021,Pancholi2021}. This is particularly the case {\it in vivo}, as placing other types of probes is far more challenging. Importantly, since these tracked proteins are typically actively interacting with the surrounding, they cannot be used as a proxy for microrheology to extract the viscoelasticity of the sample via the GSER. Indeed, GSER assumes that the probes are passive and do not interact with the environment. }

\subsubsection{Droplet Coalescence} 

In droplet coalescence assays, condensates are imaged in fluorescence or brightfield mode, and fusion events recorded. When two droplets meet they will, if liquid-like, form a neck between them, in turn transforming into an elongated shape that will eventually relax to a round drop due to surface tension (see Fig.~\ref{fig:panel}c). The higher the surface tension, the faster an elongated droplet will relax to a spherical shape; this relaxation is opposed by viscosity which will in fact resist against the re-shaping and increase the overall relaxation time. The evolving condensate is imaged at fast temporal resolution and its major and minor axis extracted via fitting of an oval shape or an asymmetric Gaussian. The ratio of minor ($l$) and major ($L$) axis yield the droplet aspect ratio $AR$ which typically display a simple exponential decay from the value of two round droplets stuck to each other to that of a single round droplet (see Fig.~\ref{fig:panel}c)
\begin{equation}
AR(t)=\frac{L(t)}{l(t)}=1+(AR_0)e^{-t/\tau} \, ,
\label{eq:fusion}
\end{equation}
where $\tau$ is the relaxation time and it is proportional to characteristic droplet size $\bar{L}$ (often taken as the average of droplets' diameters at $t=0$, i.e.  $\bar{L} = \sqrt{(L_0-l_0)l_0}$):
\begin{equation}
\tau = \frac{\gamma}{\eta} \bar{L}  \, .
\label{eq:fusion_time}
\end{equation}
By fitting the relaxation times obtained by tracking the evolution of the aspect ratio for droplets of different sizes one thus expects a straight line with slope $\eta/\gamma$, the so-called capillary velocity (the ratio between the viscosity $\eta$ and the surface tension $\gamma$)~\cite{Brangwynne2009,Brangwynne2011,Alshareedah2021b}. Similarly to FRAP, this technique is straightforward and does not require heavy post-processing. The downside is that it is not possible to extract the values of $\eta$ and $\gamma$ separately and other techniques should be coupled to it, e.g. microrheology~\cite{Elbaum-Garfinkle2015b,Alshareedah2021b}.

Interestingly, the coalescence of droplets is one of two mechanisms expected for the growth of liquid condensates. Alongside coalescence due to fusion of neighbouring droplets, the so-called ``Ostwald ripening'' process predicts that smaller droplets should shrink at the benefit of the growth of bigger ones due to energy minimisation and surface tension~\cite{chaikin2000principles,Rosowski2020,michieletto2019role}. Both mechanisms predict that the typical size of the droplets should grow in time as $r \sim t^{1/3}$~\cite{chaikin2000principles}. On the other hand, it was recently found that {\it in vivo} there is (i) no evidence of Ostwald ripening and (ii) the growth of nuclear droplets follows a scaling law $r(t) \sim t^{0.12}$ slower than that expected for either Ostwald ripening or coalescence~\cite{Lee2021}. These two unexpected observations were explained considering that {\it in vivo} nuclear condensates grow in a viscoelastic medium, i.e. the chromatin, which is more akin to a melt of polymers~\cite{Halverson2013} than to a purely viscous fluid. The behaviour of liquid-liquid phase separation within viscoelastic environments is rather unexplored and only recently started to be addressed in {\it in vitro} experiments~\cite{Style2018,Fernandez-Rico2022} and theory~\cite{Rosowski2020}. 

Finally, it is worth noting that experiments measuring passive droplet coalescence have a number of potential pitfalls. They may be affected by the surface onto which the droplets sit and diffuse, e.g. if the surface is not perfectly hydrophobic, it will create an effective friction that will slow down the fusion dynamics. It also relies on stochastic events, and thus is typically inefficient and time consuming. Finally, for droplets with large surface tension, the fusion event may be very fast thereby rendering the analysis challenging. To circumvent these problems, droplets were recently forced to fuse by using optical tweezers~\cite{Alshareedah2021b,Kaur2019,Patel2015a,Wang2018,Ghosh2020}. This technique removes surface effects because the droplets are trapped in the bulk before sedimentation and it provides a finer control over the timescales of fusion and number of events recorded~\cite{Alshareedah2021b}. The analysis is typically done on the recorded laser signal and considering an exponential relaxation process (droplet relaxation) coupled to the linear movement (constant speed) of the trap across the droplet as follows~\cite{Alshareedah2021a}
\begin{equation}
    S(t) = a e^{-t/\tau} + b t + c
\end{equation}
where $a$, $b$, $c$ and $\tau$ are fitting parameters.

It should be noted that the analysis of the passive and active droplet coalescence described above assumes Newtonian droplets which display a single relaxation timescale associated with their viscosity and surface tension (see Eq.~\eqref{eq:fusion_time}). In the case of viscoelastic droplets the analysis is less straightforward because the droplets may themselves display one, or multiple, relaxation timescales associated to the transition from elastic to viscous behaviour~\cite{Ghosh2020,Ghosh2021,Alshareedah2021c}. For instance, homotypic viscoelastic droplets pushed together by optical traps are not expected to be able to fuse until after the internal relaxation timescale~\cite{Ghosh2021} and in this sense their apparent viscosity measured using the Newtonian approximation should yield much higher values than the real one. At the same time, heterotypic droplets made of different polymers would require specific experimental conditions (e.g., temperature, pH, osmolarity) to favour miscibility (mixing) of the polymer species. From Flory-Huggins theory, the longer the polymers the lower the critical Flory parameter needed to trigger phase separation in heterotypic polymer solutions as $\chi_c \sim 1/N$, with $N$ the length of the polymers~\cite{Gennes1979}. When outside the miscibility region, heterotypic droplets are still seen to display a range of arrangements (wetting, partial engulfment or complete engulfment) depending on the relative surface tensions~\cite{Feric2016,Kaur2019}. 

\dmi{Finally, we mention that droplet fusion has been extensively used in the literature as evidence of the liquid nature of the protein condensates. Passive droplet coalescence is in fact particularly suited {\it in vivo} as it can be visualised with standard fluorescence or confocal microscopes~\cite{Kroschwald2015}, although tracking the aspect ratio of sub-micron size droplets in vivo is very challenging~\cite{Sabari2018a,Gopal2017}. On the contrary, active droplet coalescence has only been done {\it in vitro} as it requires a significant difference in refractive index between the inside and outside of the droplets in order to trap them.}

\subsection*{Models of Phase Separation}
Currently, liquid-liquid phase separation (LLPS) is the most invoked mechanism to explain the appearance of membraneless droplets. In spite of this, recent works cast some doubts on the validity of LLPS to explain some experimental observations~\cite{Peng2019a,McSwiggen2019}, such as the wide difference in the FRAP recovery for different components of heterotypic condensates~\cite{Keenen2021} or the unrestricted diffusion of PolII across replication compartments~\cite{McSwiggen2019a}.  
An alternative model to LLPS that may explain these findings is the so-called ``bridging induced attraction'', or ``bridging induced phase separation'' (BIPS) model~\cite{Brackley2013,Brackley2017,Brackey2020,Ryu2021} (also referred to as ``polymer-polymer phase separation'' in some works~\cite{Erdel2018a}). 

BIPS is a demixing process qualitatively different from LLPS in that it requires (i) a long polymeric substrate and (ii) proteins with two or more binding sites, i.e. multivalent. As we describe in detail in this section, BIPS ultimately yields condensates reminiscent of associating polymers which are well-known to display viscoelastic behaviours~\cite{Rubinstein2001}. Typical examples of BIPS are the condensation {\it in vitro} of DNA with HP1~\cite{Keenen2021}, H1~\cite{Muzzopappa2021} and yeast cohesin~\cite{Ryu2021}. Because of the multivalent binding (or effective multivalent binding due to di/oligo-merisation) these proteins are able to form loops on the polymer and locally increase the concentration of available binding sites. The nucleation of such a polymer loop triggers a positive feedback, as a locally larger concentration of binding sites will attract more proteins, which will in turn form more loops and themselves increase the density of polymer segments, or binding sites. Additionally, there is an entropic push to cluster proteins together therefore creating fewer distinct loops within the same polymer~\cite{Brackley2013,Brackley2017,Brackey2020}. This entropic clustering mechanism was first proposed in Ref.~\cite{Brackley2013}. BIPS ultimately drives the coarsening and coalescence of clusters of proteins interwoven within the polymer (DNA or chromatin) substrate~\cite{Brackley2013,Brackley2017}. Importantly, these protein clusters -- which would not exist without the polymeric substrate -- appear to have surface tension and to grow as expected for classic Ostwald ripening and coalescence of demixing liquid droplets, but could also be arrested by introducing non-equilibrium binding modes~\cite{Brackley2017,Michieletto2019prl} or by limiting their binding to specific DNA sites~\cite{Brackley2013,Brackley2016nar}. Simulated FRAP on BIPS-driven protein clusters show that they can recover by exchanging with the soluble pool while the underlying polymeric framework is glassy, and evolves on much longer timescales~\cite{Brackley2017}, in line with what observed in FRAP experiments of HP1$\alpha$ and DNA droplets~\cite{Keenen2021}. 
BIPS also predicts that FRAP recovery curves for the polymer itself would depend on its length, as the longer the polymer the more entangled and the slower to rearrange~\cite{Brackley2017}, consistent with the observations in Ref.~\cite{Muzzopappa2021}. 

BIPS can also explain the recent observation of enrichment of PolII into viral replication compartments (RC) in Herpex Simplex infected cells~\cite{McSwiggen2019a}. Here, BIPS is triggered by the fact that the viral genome is enriched in highly accessible (ATAC) sites, and so it displays a large number of non-specific binding sites for PolII in turn triggering the feedback loop described above. This is consistent with the single molecule tracking data suggesting weak transient binding events of PolII inside the RC~\cite{McSwiggen2019a}. A similar argument explains the unrestricted motion of PolII in and out of RC due to the abundance of non-specific binding sites inside the RC. Furthermore, in general, BIPS-driven droplets are far less sensitive on the concentration of the protein component: in LLPS, a certain critical concentration must be attained before triggering demixing. Increasing the concentation of protein past this critical concentration will not change the concentration within the dense phase, only its volume. This is not the case in BIPS, which can be triggered at very low protein concentrations and can increase the concentration of the dense phase as long as it displays free binding sites. Another peculiar feature of BIPS is that the polymer is compacted by the presence of the proteins, yet this process does not necessarily produce a dense and inaccessible polymer globule, on the contrary the degree of polymer compaction depends on the number of binding sites in the protein of interest; for instance, yeast cohesin was found to maintain a rather open DNA structure compatible with bi-valent binding~\cite{Ryu2021}.

How can we distinguish LLPS from BIPS using the methods described above? First, due to the presence of a long polymeric substrate, BIPS-driven condensates are typically viscoelastic and reminiscent of systems of associating polymers, yet their protein component may display fast exchange within the condensate and with the soluble pool, as in the case of DNA and HP1$\alpha$ condensates~\cite{Keenen2021}. Thus, microrheology with large probes, FRAP on the polymeric component and droplet coalescence have the best chances to reveal the elastic contribution to the droplet material properties. At the same time, while LLPS-driven droplets can be either viscous or viscoelastic, BIPS-driven droplets cannot be purely viscous if triggered by a long polymer substrate. Similarly, active microrheology and active/passive droplet coalescence should reveal their sluggishness due to the intrinsically slow polymer network, which can be thought of as a transiently cross-linked polymer gel. The intrinsic relaxation time of the droplet is thus related to the timescales of the cross-links, i.e. protein on/off kinetics~\cite{Brackley2017} and stoichiometry~\cite{Ronceray2022,Khanna2019}, and entanglements, related to the length of the polymer~\cite{Rubinsteina}. In line with this, in Refs.~\cite{Keenen2021} and ~\cite{Muzzopappa2021}, droplets made with longer DNA presented more irregular and non-spherical shapes with DNA-length dependent FRAP recovery. To the best of our knowledge no microrheology was performed on those droplets but we expect them to display a strong elastic component that grows with the length of the polymer substrate. 

Distinguishing LLPS from BIPS {\it in vivo} is more challenging as microrheology cannot be easily performed in the cell nucleus. For a protein known or suspected to bind DNA/chromatin, then FRAP or FCS should be performed on both, the protein and the underlying DNA/chromatin component (or on fluorescently labelled histones as a proxy for chromatin). Additionally, since BIPS cannot be triggered without a long polymeric substrate, purifying the protein of interest and testing its tendency to phase separate in presence/absence of long DNA {\it in vitro} is potentially the best way to test LLPS versus BIPS.    

\dmii{We note that LLPS and BIPS are not the only two models that have been proposed to explain the behavoiur of intrinsically disordered proteins. Indeed, the ``Phase-separation-aided bond percolation'' (PSBP) model~\cite{Choi2020a} is an appealing model for condensates made of multivalent proteins. It connects the condensate material properties to critical percolation phenomena in systems of associating polymers, and is particularly appropriate to explain ageing and hardening of condensates, as seen in Ref.~\cite{Jawerth2020}. PSBP differs from LLPS as it introduces complexity in the form of long-lived bridging between proteins. PSBP also differs from BIPS in that the latter does not require protein-protein attraction but it requires a long polymer substrate (such as long DNA or RNA segments) and multivalent DNA/RNA binding proteins to initiate and form condensates, as seen in Ref.~\cite{Ryu2021} for yeast cohesin. In turn, the viscoelastic nature of the condensates largely relies on the entanglement and transient cross-linking of the long scaffold, rather than the protein-protein associativity. Note that LLPS may, in some cases, act as a precursor of PSBP and we therefore stress that the most glaring difference between LLPS/PSBP and BIPS is the presence of an underlying polymer scaffold to which proteins bind to.
}

\dmii{We argue that, as often is the case in biology, multiple mechanisms may be at play and may be in place to address different biological requirements. For instance, while liquid condensates may accelerate reactions, viscoeleastic condensates may offer transient structural support~\cite{Michieletto2019review}. In turn, terminology such as LLPS should not be used without appropriate evidence~\cite{McSwiggen2019} and other mechanisms such as BIPS and PSBP should be considered.}

\section*{Discussion and Conclusions}
While some may argue that protein condensation in biology is ``just a phase'', we feel that this field is putting the spotlight on previously underappreciated universal mechanisms that are employed by life to sense, respond, organise and control intra-cellular processes. Generic physical principles such as phase separation and demixing which were traditionally studied by physicists, chemists and engineers to describe metal alloys and binary fluids~\cite{Gennes1979,Cahn1959a} now find application to describe the behaviour of proteins and nucleic acids inside living cells. This clearly creates a unique nexus and a meeting point between very different disciplines. In the near future this field will likely attract an even broader interdisciplinary audience. 

\dmi{The main point of this review is that experiments performed in the last decade have uncovered that phase separation (LLPS or BIPS) -- thermodynamically-driven, reversible demixing of liquid phases --  can trigger the onset of non-Newtonian fluid behaviours, e.g. viscoelasticity, via the interaction of intrinsically disordered protein domains with themselves or with nucleic acids in the dense phase. Because of this, LLPS/BIPS and viscoelasticity should in fact be thought of as two sides of the same coin. The observation that condensates are round and that they coalesce over long times is not unambiguous proof that they are simple viscous liquids. Additionally, even if they were simple liquids at early times, there is no guarantee {\it a priori} that they will always remain so~\cite{Jawerth2020}. Among the techniques we reviewed above, active and passive microrheology stand out, as they can provide quantitative information on the full viscoelastic spectrum of the condensates~\cite{Jawerth2018,Jawerth2020,Alshareedah2021c,Ghosh2021}. Microrheology will in fact be able to assess if a protein droplet is a simple liquid (no elastic modulus $G^\prime$) or if it displays elastic response to deformations and at which timescales they appear to be dominant over the viscous ones. In spite of this, microrheology is very challenging to perform {\it in vivo} and it is not a familiar technique in biology (yet). More commonly employed techniques are FRAP, FCS and droplet coalescence assays which can probe the viscosity and surface tension of the condensates, but cannot quantify the response of the droplets to deformations occurring at different frequencies. Additionally, FRAP and FCS are ``nanorheology'' techniques as they rely on small probes (GFP or fluorescent proteins) which are likely to underestimate the bulk viscosity of the sample and fail to detect its elasticity. }  

\dmi{It is also appropriate to mention here that the inside of the cell is a busy environment where many chemicals and components are involved in energy-consuming processes. It should therefore not be a surprise if certain protein condensates were controlled by out-of-equilibrium processes, for instance involving post-translational modifications~\cite{Michieletto2019prl,Coli2019,Hilbert2021,Narayanan2019,Eeftens2021}, reaction-diffusion networks~\cite{Zwicker2018,Zwicker2014,Zwicker2015}, or environment elasticity~\cite{Style2018,Shin2018,Lee2021}. In these cases, it would be even more intuitive to expect unconventional flow properties associated with the non-equilibrium nature of the droplets.}

\dmi{While we refrain from discussing in detail about simulations of protein phase separation (see Ref.~\cite{Posey2018} for a comprehensive review) we mention that, currently, simulations are mainly concerned to capture the demixing and phase behaviours rather than the viscoelastic properties of droplets~\cite{Dignon2018a,Dignon2019,Dignon2018,Ronceray2022}. The reason for this may be that protein condensation encompasses a broad range of time- and length-scales~\cite{Brangwynne2015,Choi2020,Joseph2021a} which are difficult to capture within the same model. For instance, near atomistic models may be needed to capture the correct phase behaviour, while more coarse grained models are necessary to model bulk viscoelasticity. Because of this, multi-scale modelling of protein condensates and their viscoelastic properties is a field that is just beginning to appear and will likely attract a number of computational researchers from soft matter, rheology, polymer physics and fluid mechanics.}

The next steps in this quickly expanding field will certainly involve more research {\it in vivo}; the connection between the condensation of a certain protein {\it in vitro} -- sometimes under extreme crowding or salt conditions -- and its biological relevance is oftentimes weak or circumstantial. Alongside this, we feel that often condensates are mis-classified as originating from liquid-liquid phase separation because too little is known about other potential mechanisms~\cite{McSwiggen2019}. As discussed above, bridging-induced phase separation~\cite{Brackley2013,Brackley2017,Ryu2021} is an appealing alternative mechanisms to explain a range of observations, such as viral replication compartments in herpex simplex infected cells~\cite{McSwiggen2019a} or even Polycomb~\cite{Eeftens2021} and heterochromatin~\cite{Brackley2016nar,Michieletto2016prx,DiPierro2016} compartments. We expect that in the DNA-rich eukaryotic nucleus, this mechanism may well dominate over more traditional liquid-liquid phase separation. 

Finally, we stress that the non-trivial flow behaviours of certain protein condensates may have important biological relevance. For instance, viscous droplets may provide a crucible to accelerate reactions or sequester reagents, while condensates with elastic components may provide structural support to shape genome organisation~\cite{Michieletto2019rnareview,Nozawa2017}, control chromatin dynamics~\cite{Khanna2019} and regulate enhancer-promoter interactions~\cite{Sabari2018}. By using the methods and the concepts provided in this review, we thus hope that the research community will be better equipped at answering these outstanding questions.

\begin{acknowledgement}
DM is a Royal Society University Research Fellow. This project has received funding from the European Research Council (ERC) under the European Union's Horizon 2020 research and innovation programme (grant agreement No 947918, TAP). MM is an MRC funded Cross-Disciplinary Post-Doctoral Fellow (MC UU 00009/2). We thank the members of the TAPLab for the critical reading of the manuscript and useful comments.
\end{acknowledgement}


%
\bibliography{library,library2}

\end{document}